\documentclass[12pt]{iopart}

\usepackage{amssymb}
\usepackage{iopams}
\usepackage{graphicx}
\usepackage{cite}
\usepackage{multirow}
\usepackage{dcolumn}% Align table columns on decimal point
\usepackage{bm}% bold math
\usepackage{makecell}

\begin{document}

\title[]{Study of baryon octet electromagnetic form factors in perturbative chiral quark model}

\author{X. Y. Liu$^{1,2}$, K. Khosonthongkee$^{1,2}$, A. Limphirat$^{1,2}$ and Y. Yan$^{1,2}$}

\address{$^1$ School of Physics, Institute of Science, Suranaree University of Technology, Nakhon Ratchasima 30000, Thailand}
\address{$^2$ Thailand Center of Excellence in Physics (ThEP), Commission on Higher Education, Bangkok 10400, Thailand}

\ead{lxy\_gzu2005@126.com,yupeng@sut.ac.th}

\begin{abstract}
The electromagnetic properties of baryon octet are studied in the perturbative chiral quark model~(PCQM). The relativistic quark wave function is extracted by fitting the theoretical results of the proton charge form factor to experimental data and the predetermined quark wave function is applied to study the electromagnetic form factors of other octet baryons as well as magnetic moments, charge and magnetic radii. The PCQM results are found, based on the predetermined quark wave function, in good agreement with experimental data.
\end{abstract}

\submitto{\jpg} % Uncomment for Submitted to journal title message
%\maketitle % Comment out if separate title page not required

\section{Introduction}{\label{sec:Intro}

The perturbative chiral quark model (PCQM) was developed a decade ago~\cite{Lyubovitskij1:2001,Lyubovitskij2:2002} and has become one of the most successful
approaches in the low energy particle physics. In the PCQM, baryons are considered as the bound states of three relativistic valence quarks while a cloud of pseudoscalar mesons, as the sea-quark excitations, is introduced for chiral symmetry requirements. The quarks move in a self-consistent field, represented by scalar $S(r)$ and vector $V(r)$ components of a static potential providing confinement while the interactions between quarks and mesons are achieved by the nonlinear $\sigma$ model in the PCQM. The PCQM has been successfully applied to the electromagnetic and axial form factors of baryons~\cite{Lyubovitskij3:2001,Cheedket4:2004,Khosonthongkee10:2004,Faessler5:2008}, low-energy meson-baryon scatterings~\cite{Lyubovitskij6:2001}, electromagnetic excitations of nucleon resonances~\cite{Pumsa-ard7:2003}, nucleon polarizabilities~\cite{Dong8:2006}, neutron electric dipole form factor~\cite{Dib9:2006}, etc. These studies demonstrate the PCQM is one of the effective models in the low energy hadron physics, and indicate that the virtual meson cloud exists mainly outside of the quark core.

In recent years the electroweak form factors have been studied in chiral perturbation theory~\cite{Kubis46:2001,Schindler47:2001}, various relativistic quark models~\cite{Inanov51:1996,Franklin45:2003,Faessler52:2006,Faessler53:2006,Cloet54:2009,Ramalho41:2011,Gutche55:2012,Ramalho42:2013}, Lattice-QCD~\cite{Wang50:2009,Wang48:2009,Lin43:2009,Lin49:2009}, etc. in which the theoretical
results are comparable with experimental data. But the PCQM theoretical results~\cite{Lyubovitskij3:2001,Cheedket4:2004,Khosonthongkee10:2004,Faessler5:2008} of the electromagnetic and axial form factors of baryons are in good agreement with experimental data only at very low momentum transfer, descending quickly with the momentum transfer increasing. In those works~\cite{Lyubovitskij3:2001,Cheedket4:2004,Khosonthongkee10:2004,Faessler5:2008}, a variational Gaussian ansatz has been employed for the quark wave function. One may argue that it is the Gaussian-type quark wave function of baryons which leads to the theoretical predictions for the form factors of baryons consistent well with experimental data only at very low momentum transfer. In this work we extract the quark wave function by fitting the PCQM theoretical result of the proton charge form factor to experimental data, and study the electromagnetic form factors of octet baryons with the predetermined wave function.

The paper is organized as follows. In~\sref{sec:PCQM&EMFF} we introduce the PCQM and present the theoretical expressions of octet baryon electromagnetic form factors in the framework of the PCQM. The quark wave function is extracted by fitting the theoretical result of the proton charge form factor to experimental data in~\sref{sec:QWF}. The numerical results based on the predetermined quark wave function and discussion are given in~\sref{sec:Results}.

\section{Electromagnetic form factors in the PCQM}\label{sec:PCQM&EMFF}

The PCQM~\cite{Lyubovitskij1:2001,Lyubovitskij2:2002} is based on an effective chiral Lagrangian describing baryons by a core of the three valence quarks, moving in a central Dirac field with $V_{\textrm{eff}}(r)=S(r)+\gamma_0V(r)$, where $r=|\vec x|$. In order to respect chiral symmetry, a cloud of Goldstone bosons ($\pi$, $K$~and $\eta$) is included as small fluctuations around the three-quark core in SU(3) extension. With an unitary chiral rotation, as shown in Refs.~\cite{Lyubovitskij6:2001,Khosonthongkee10:2004}, the Weinberg-type Lagrangian of the PCQM is derived,
\begin{eqnarray}
\fl\mathcal{L}^W(x)= \mathcal{L}_0(x)+\mathcal{L}^W_I(x)+o(\vec{\pi}),\label{eq:WT-Lagrangian}\\
\fl\mathcal{L}_0(x)= \bar{\psi}(x)\big[i\partial\!\!\!/-\gamma^{0}V(r)-S(r)\big]\psi(x)
-\frac{1}{2}\Phi_i(x)\big(\Box+M_{\Phi}^2 \big) \Phi^i(x),\label{WT-0}\\
\fl\mathcal{L}_I^W(x)=\frac{1}{2F}\partial_\mu\Phi_i(x)\bar{\psi}(x)\gamma^\mu\gamma^5
\lambda^i\psi(x)+\frac{f_{ijk}}{4F^2}\Phi_i(x)\partial_\mu\Phi_j(x)
\bar\psi(x)\gamma^\mu\lambda_k\psi(x)\label{eq:WT-int},
\end{eqnarray}
where $f_{ijk}$ are the totally antisymmetric structure constant of $SU(3)$, the pion decay constant $F=88$ MeV in the chiral limit, $\Phi_i$ are the octet meson fields,
and $\psi(x)$ is the triplet of the $u$, $d$, and $s$ quark fields taking the form
\begin{equation}
\psi(x) =\left(
                                                 \begin{array}{c}
                                                  u(x) \\
                                                  d(x) \\
                                                  s(x) \\
                                                 \end{array}
                                           \right).
\end{equation}
The quark field $\psi(x)$ could be expanded in
\begin{equation}
\psi(x)=\sum_\alpha\left(b_\alpha u_\alpha(\vec{x})\,e^{-i\mathcal{E}_\alpha t}+d^\dagger_\alpha\upsilon_\alpha (\vec{x})e^{i\mathcal{E}_\alpha t}\right),
\end{equation}
where $b_\alpha$ and $d^\dag_\alpha$ are the single quark
annihilation and antiquark creation operators. The ground state quark wave function $u_0(\vec x)$ may, in general, be expressed as
\begin{equation}
u_0(\vec{x})=\left(\begin{array}{c}g(r)\\\large{i\vec{\sigma}\cdot\hat{x}f(r)}
\end{array}\right)\chi_s\chi_f\chi_c,
\end{equation}
where $\chi_s$, $\chi_f$ and $\chi_c$ are the spin, flavor and color quark wave functions, respectively.

The calculation technique in the PCQM is based on the Gell-Mann and Low theorem~\cite{Gell-Mann42:1951}, in which the expectation value of an operator $\hat O$ can be calculated from
\begin{eqnarray}\label{eq:Gell-Mann-Low}
\fl\langle\hat{O} \rangle=\,^B\langle \phi_0|\sum_{n=0}^\infty \frac{i^n}{n!}\int d^4x_1\cdots\int d^4x_nT[\mathcal{L}_I^W(x_1)\cdots \mathcal{L}_I^W(x_n)\hat O]|\phi_0\rangle^B_c,
\end{eqnarray}
where the state vector $|\phi_0\rangle^B$ corresponds to the unperturbed three-quark states projected onto the respective baryon states, which are constructed in the
framework of the $SU(6)$ spin-flavor and $SU(3)$ color symmetry. The subscript \textit{c} in~\eref{eq:Gell-Mann-Low} refers to contributions
from connected graphs only. $\mathcal{L}_I^W(x)$ is the quark-meson interaction Lagrangian as given in~\eref{eq:WT-int}. In the framework of the PCQM, the charge and magnetic form factors of octet baryons in the Breit frame are defined by
\begin{eqnarray} \label{Charge-FF}
\chi^\dag_{B_s'} \chi_{B_s}G^B_E(Q^2)&=&\,^B \langle{\phi_0}|\sum_{n=0}^n\frac{i^n}{n!}
\int\delta(t)d^4xd^4x_1\cdots d^4x_n e^{-i q\cdot
x}\nonumber \\
&&\times T[\mathcal{L}_I^W(x_1)\cdots
\mathcal{L}_I^W(x_n)j^0(x)]|\phi_0\rangle_c^B,
\end{eqnarray}
\begin{eqnarray} \label{Magnetic-FF}
\chi^\dag_{B_s'} \frac{i\vec{\sigma}\times\vec{q}}{m_B+m'_B}\chi_{B_s}G^B_M(Q^2)
&=&\,^B\langle{\phi_0}|\sum_{n=0}^n\frac{i^n}{n!}
\int\delta(t)d^4xd^4x_1\cdots d^4x_ne^{-iq\cdot x}\nonumber \\
& &\times T[\mathcal{L}_I^W(x_1)\cdots
\mathcal{L}_I^W(x_n)\vec{j}(x)]|\phi_0\rangle_c^B.
\end{eqnarray}
Here, $G_E^B(Q^2)$ and $G_M^B(Q^2)$ are the charge and magnetic form factors of octet baryons with the space-like squared momentum transfer $Q^2$, which is carried out by the electromagnetic current. In the Breit frame, the initial momentum of the baryons is $p=(E,-\vec q/2)$, the final momentum is $p=(E,\vec q/2)$, and the four-momentum of the photon is $q=(0, \vec q)$. Thus, $Q^2=-q^2={\vec q}\,^2$. $m_B$ is the mass of baryons. $\chi_{B_s}$ and $\chi_{B_{s'}}^\dag$ are the baryon spin wave functions in the initial and final states, $\vec \sigma$ is the baryon spin operator, and $j^\mu$ is the electromagnetic current
\begin{equation}
j^\mu=j^\mu_\psi+j^\mu_\Phi+j^\mu_{\psi\Phi}+\delta j^\mu_\psi,
\end{equation}
which contains the quark current $j^\mu_\psi$, the charged pseudoscalar mesons current $j^\mu_\Phi$, the quark-meson coupling current $j^\mu_{\psi\Phi}$, and $\delta j^\mu_\psi$, a current arising from the counterterm. The currents in the above equation take the forms,
\begin{eqnarray}
j^\mu_\psi & = & \bar\psi\gamma^\mu {\cal Q}\psi,\label{eq:q-current}\\
j^\mu_\Phi & = & \bigg[f_{3ij}+\frac{f_{8ij}}{\sqrt{3}}\bigg]
\Phi_i\partial^\mu\Phi_j\label{eq:m-current}\\
j^\mu_{\psi\Phi} & = & \bigg[f_{3ij}+\frac{f_{8ij}}{\sqrt{3}}\bigg]\frac{\Phi_j}{2F}
\bar\psi\gamma^\mu\gamma^5\lambda_i\psi,\label{eq:qm-current}\\
\delta j^\mu_\psi & = & \bar\psi\big(Z-1\big)\gamma^\mu  {\cal Q}\psi,\label{eq:c-current}
\end{eqnarray}
where $ {\cal Q}$ is the quark charge matrix $ {\cal Q}=\textrm{diag}\{2/3, -1/3, -1/3\}$, and the renormalization constants $\hat Z$ and $\hat Z_s$ are defined as
\begin{eqnarray}
\hat Z=1-\frac{3}{4(2\pi F)^2}\int_0^\infty dk k^4 F_{I}^2(k^2)\bigg[\frac{1}{\omega_\pi^3(k^2)}+\frac{2}{3\omega_K^3(k^2)}
+\frac{1}{9\omega_\eta^3(k^2)}\bigg],\\
\hat Z_s=1-\frac{1}{(2\pi F)^2}\int_0^\infty dk k^4 F_{I}^2(k^2)\bigg[\frac{1}{\omega_K^3(k^2)}
+\frac{1}{3\omega_\eta^3(k^2)}\bigg],
\end{eqnarray}
with $\omega_\Phi(k^2)=\sqrt{M_\Phi^2+k^2}$ and the vertex function $F_{I}(k)$ for the $qq\Phi$ system taking the form
\begin{eqnarray}
F_{I}(k)=2\pi \int_0^\infty dr \int_0^\pi d\theta r^2 \sin\theta e^{i k r \cos\theta}[g(r)^2+f(r)^2\cos2\theta].
\end{eqnarray}

\begin{figure}
\begin{center}
\includegraphics[width=7cm,height=11.5cm]{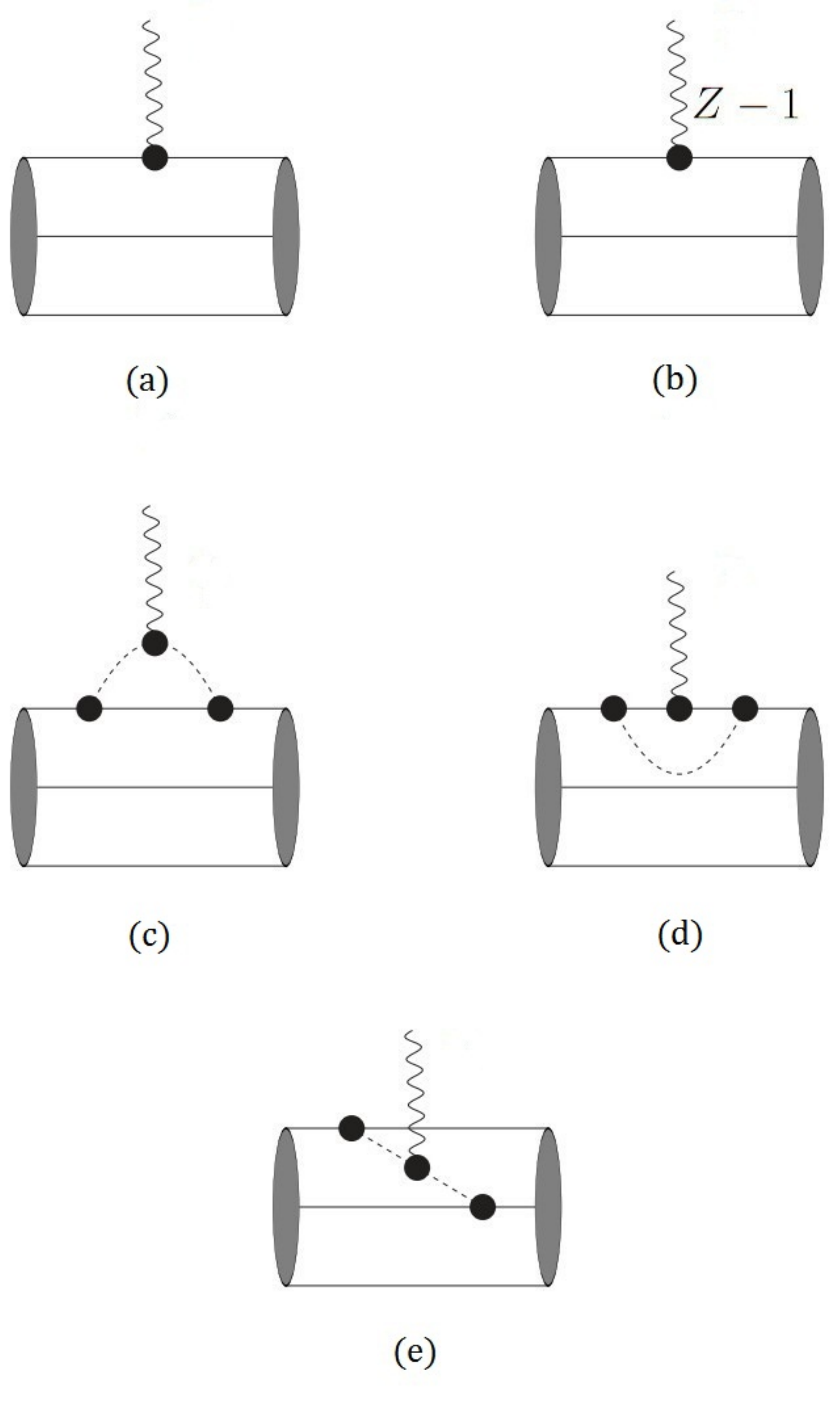}
\caption{\label{fig:EMFF}Diagrams contributing to the electromagnetic form factors of the baryon octet: three-quark diagram (a), three-quark counterterm diagram (b), meson cloud diagram (c), vertex correction diagram (d), and meson-in-flight diagram (e).}
\end{center}
\end{figure}
In accordance with the interaction Lagrangian $\mathcal{L}_I^W(x)$ in~\eref{eq:WT-int} and the electromagnetic current $j^\mu$ in~\eref{eq:q-current}-\eref{eq:c-current}, there are five Feynman diagrams, as shown in~\fref{fig:EMFF}, contributing to the electromagnetic form factors to the one-loop order~\cite{Lyubovitskij3:2001,Cheedket4:2004}. The contributions of these diagrams are derived as follows:\\
{\bf(a)} Three-quark core leading-order diagram~(LO)
\begin{eqnarray}
G_E^B(Q^2)\big|_{LO}&=& a_1^B G_E^p(Q^2)\big|_{LO}^{3q},\\
G_M^B(Q^2)\big|_{LO}&=& b_1^B \frac{m_B}{m_N}G_M^p(Q^2)\big|_{LO}^{3q},
\end{eqnarray}
where
\begin{eqnarray}
G_E^p(Q^2)\big|_{LO}^{3q}=2\pi\int_0^\infty dr \int_0^\pi d\theta r^2 \sin\theta[g(r)^2+f(r)^2] e^{i Q r\cos\theta},\\
G_M^p(Q^2)\big|_{LO}^{3q}=\frac{4\pi i m_N}{Q}\int_0^\infty dr \int_0^\pi d\theta r^2 \sin(2\theta) g(r)f(r) e^{i Q r\cos\theta}.
\end{eqnarray}
{\bf (b)} Three-quark core counterterm diagram~(CT)
\begin{eqnarray}
G_E^B(Q^2)\big|_{CT}^{3q}=\big[a_2^B(\hat Z-1)+a_3^B(\hat Z_s-1)\big]G_E^p(Q^2)\big|_{LO}^{3q},\\
G_M^B(Q^2)\big|_{CT}^{3q}=\big[b_2^B(\hat Z-1)+b_3^B(\hat Z_s-1)\big]\frac{m_B}{m_N}G_M^p(Q^2)\big|_{LO}^{3q}.
\end{eqnarray}
{\bf (c)} Meson-cloud diagram~(MC)
\begin{eqnarray}
\fl G_E^B(Q^2)\big|_{MC}=\frac{1}{2(2\pi F)^2}\int_0^\infty dk \int_{-1}^1 dx k^2(k^2+k Q x)F_{I}(k)F_{I}(k_+) t_E^B(k^2, Q^2, x)\big|_{MC},\\
\fl G_M^B(Q^2)\big|_{MC}=\frac{5 m_B}{6(2\pi F)^2}\int_0^\infty dk k^4\int_{-1}^1 dx (1- x^2)F_{I}(k)F_{I}(k_+) t_M^B(k^2, Q^2, x)\big|_{MC},
\end{eqnarray}
where
\begin{eqnarray}
t_E^B(k^2, Q^2, x)\big|_{MC}=a_4^B C_\pi(k^2,Q^2,x)+a_5^B C_K(k^2,Q^2,x),\\
t_M^B(k^2, Q^2, x)\big|_{MC}=b_4^B D_\pi(k^2,Q^2,x)+b_5^B D_K(k^2,Q^2,x),
\end{eqnarray}
\begin{eqnarray}
C_\Phi(k^2,Q^2,x)=\frac{1}{\omega_\Phi(k^2)\omega_\Phi(k_+^2)
[\omega_\Phi(k^2)+\omega_\Phi(k_+^2)]},\\
D_\Phi(k^2,Q^2,x)=\frac{1}{\omega_\Phi^2(k^2)\omega_\Phi^2(k_+^2)},\\
k_+ = \sqrt{k^2+Q^2+2k\sqrt{Q^2}x}.
\end{eqnarray}
{\bf(d)} Vertex-correction diagram~(VC)
\begin{eqnarray}
\fl G_E^B(Q^2)\big|_{VC}=\frac{1}{2(2\pi F)^2}\int_0^\infty dk k^4F_{I}^2(k)G_E^p(Q^2)\big|_{LO}^{3q} \bigg[\frac{a_6^B}{\omega_\pi^3(k^2)}+\frac{a_7^B}
{\omega_K^3(k^2)}+\frac{a_8^B}{\omega_\eta^3(k^2)}\bigg],\\
\fl G_M^B(Q^2)\big|_{VC}=\frac{1}{2(2\pi F)^2}\int_0^\infty dk k^4F_{I}^2(k)G_M^p(Q^2)\big|_{LO}^{3q} \bigg[\frac{b_6^B}{\omega_\pi^3(k^2)}+\frac{b_7^B}
{\omega_K^3(k^2)}+\frac{b_8^B}{\omega_\eta^3(k^2)}\bigg].
\end{eqnarray}
{\bf (e)} Meson-in-flight diagram~(MF)
\begin{eqnarray}
\fl G_E^B(Q^2)\big|_{MF}\equiv0,\\
\fl G_M^B(Q^2)\big|_{MF}=\frac{m_B}{(2\pi F)^2}\int_0^\infty dk \int_{-1}^1 dx k^4(1-x^2) F_{I}(k)F_{I}(k_+) t_M^B(k^2, Q^2, x)\big|_{MF},
\end{eqnarray}
where
\begin{equation}
t_M^B(k^2, Q^2, x)\big|_{MF}=b_9^B D_\pi(k^2,Q^2,x)
+b_{10}^B D_K(k^2,Q^2,x).
\end{equation}
The constants $a_i^B$ and $b_i^B$, which depend on the spin and flavor of baryons, have been listed in Ref.~\cite{Cheedket4:2004}.

In the non-relativistic limit, the mean-square charge radius of a charged baryon is related to the baryon charge form factor as
\begin{equation}\label{eq:CChargeRadius}
\langle r^2_E\rangle^B=-\frac{6}{G_E^B(0)}\frac{d}{d Q^2} G_E^B(Q^2)\Big|_{Q^2=0}.
\end{equation}
For the neutral baryons, the mean-square charge radius is defined by
\begin{equation}\label{eq:NChargeRadius}
\langle r^2_E\rangle^B=-6\frac{d}{d Q^2} G_E^B(Q^2)\Big|_{Q^2=0}.
\end{equation}
In analogy, the mean-square magnetic radius is defined as
\begin{equation}\label{eq:MagneticRadius}
\langle r^2_M\rangle^B=-\frac{6}{G_M^B(0)}\frac{d}{d Q^2} G_M^B(Q^2)\Big|_{Q^2=0}.
\end{equation}

\section{Model quark wave function}\label{sec:QWF}
The theoretical expressions in the above section show that the charge and magnetic form factors are mainly determined by the wave function of the quark core. In the previous works~\cite{Lyubovitskij3:2001,Cheedket4:2004}, the Gaussian-type quark wave function of baryons is employed and the theoretical results for the form factors of baryons are consistent well with the experimental data at low momentum transfer $Q^2$.
Instead of assuming a certain type, we extract in this work the quark wave function by adjusting our theoretical result of the proton charge form factor to the experimental data, considering that the recent measurements of the proton charge form factor are in high precision and that only four Feynman diagrams in the PCQM contribute to the proton charge form factor. The radial quark wave functions $g(r)$ and $f(r)$, the upper and lower components in the ground state,
are expanded in the complete set of Sturmian functions $S_{nl}(r)$~\cite{Rotenberg11:2012},
\begin{eqnarray}
g(r)&=&\sum_{n}A_n\frac{S_{n0}(r)}{r},\label{eq:Upperg}\\
f(r)&=&r \sum_{n}B_n\frac{S_{n0}(r)}{r},\label{eq:Lowerf}
\end{eqnarray}
with
\begin{equation}\label{eq:Sturmain-func}
S_{nl}(r)=\left[{\frac{n!}{(n+2l+1)!}}\right]^{\frac{1}{2}}
(2br)^{l+1}e^{-br}L^{2l+1}_n(2br),
\end{equation}
where $L^{2l+1}_n(x)$ are Laguerre polynomials, and $b$ is the length parameter of Sturmian functions.
For the sake of simplicity, we restrict our calculations to $SU(2)$ flavor.
\begin{table}
\caption{\label{tab:Parameters} Expansion coefficients $A'_n$ and $B'_n$ determined by fitting the theoretical result of the proton charge form factor to the experimental data,
with the errors of the experimental data considered.}
\begin{indented}
\lineup
\item[]\begin{tabular}{@{}*{3}{c}}
\br
$n$ & $A'_n$ & $B'_n$\cr
\mr
   0 & 0.21966$\pm$0.00669 & 0.13892$\pm$0.01405\cr
   1 & \-0.00817$\pm$0.01204 & 0.02905$\pm$0.00510\cr
   2 & 0.00073$\pm$0.00107 & 0.01025$\pm$0.00115\cr
   3 & \-0.01312$\pm$0.00230 & 0.00072$\pm$0.00086\cr
   4 & \-0.00853$\pm$0.00150 & \-0.00092$\pm$0.00016\cr
\br
\end{tabular}
\end{indented}
\end{table}

\begin{figure}
\begin{center}
\includegraphics[width=7.8cm]{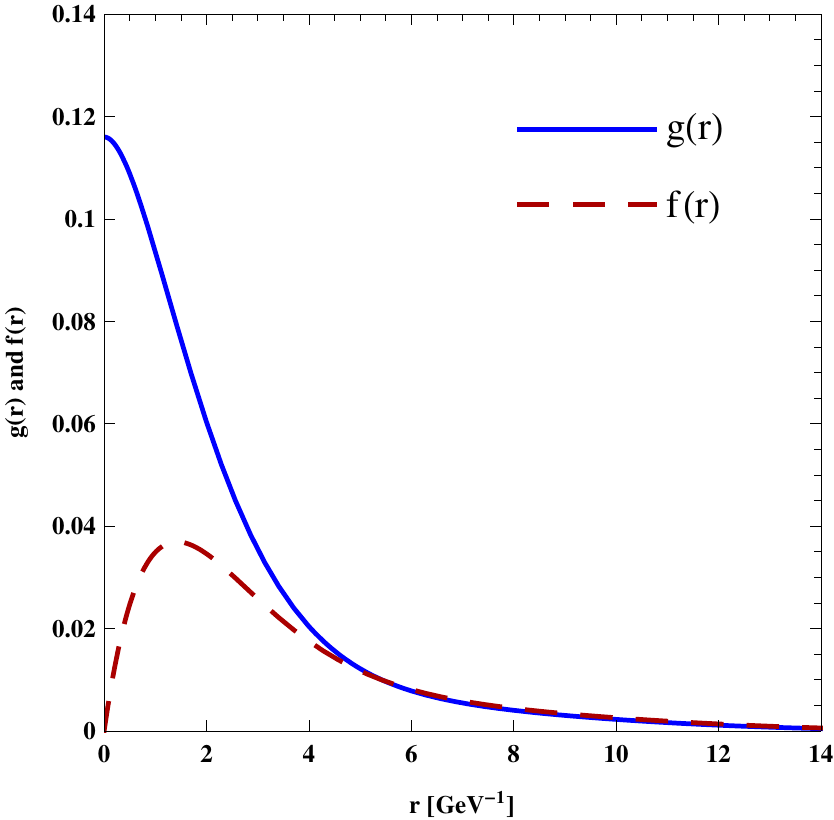}
\includegraphics[width=7.75cm]{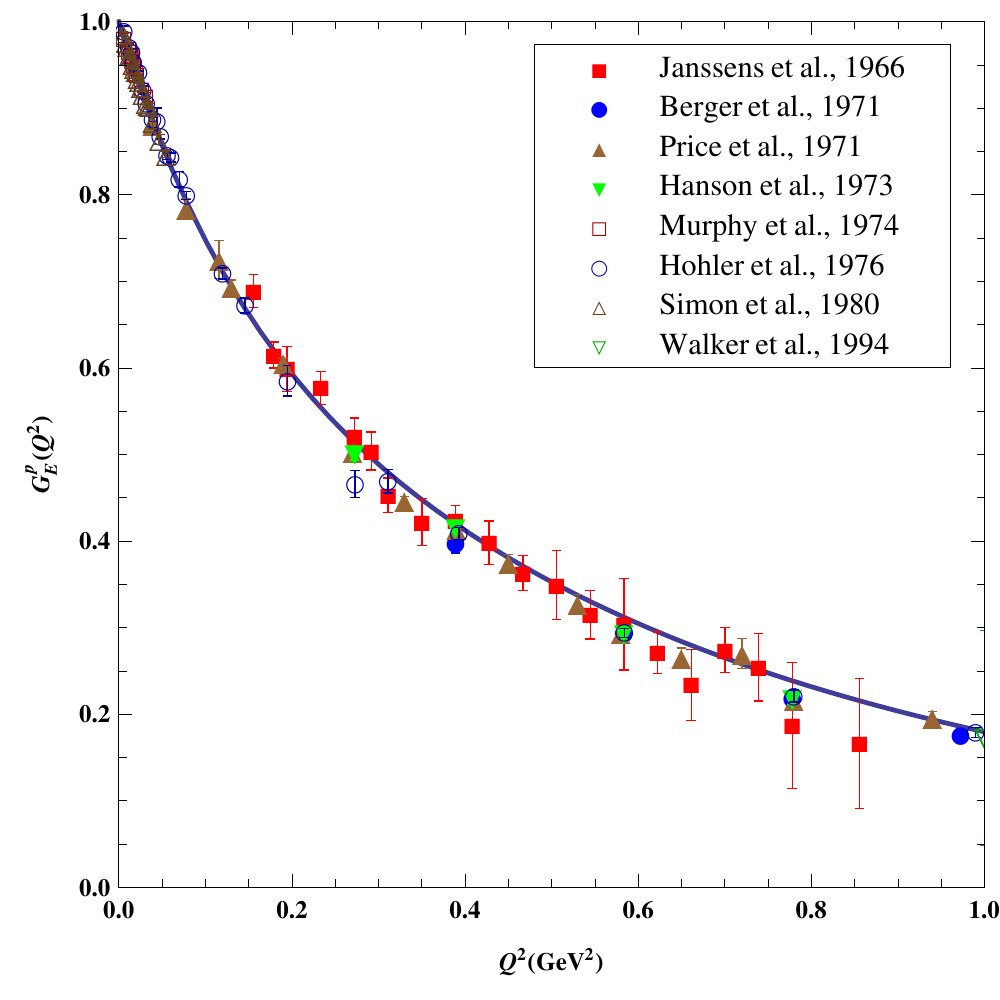}
\end{center}
\caption{Left panel: Normalized radial wave functions of the valence quarks for the upper component $g(r)$ and the lower component $f(r)$ with the central values of the expansion coefficients. Right panel: Fits of proton charge form factor to the measurements~\cite{Janssens13:1966,Berger14:1971,Price15:1971,Hanson16:1973,Murphy17:1974,Hohler18:1976,Simon19:1980,Walker20:1994}.}\label{fig:QWF}
\end{figure}

It is found that a basis of five Sturmian functions ($n=0, 1, 2, 3, 4$) is good enough to let our theoretical result of the proton charge form factor fit to the experimental data. The Sturmian function length parameter is fixed to be $b=0.5$ GeV, and the expansion coefficients $A_n$ and $B_n$ are compiled in~\tref{tab:Parameters}, where we redefine $A'_n=A_nb^{-1/2}$ and $B'_n=B_nb^{-3/2}$ to let the $A'_n$ and $B'_n$ be dimensionless. The expansion coefficients are determined by adjusting the theoretical result of proton charge form factor to the experimental
data~\cite{Janssens13:1966,Berger14:1971,Price15:1971,Hanson16:1973,Murphy17:1974,Hohler18:1976,Simon19:1980,Walker20:1994}, in which the errors of the experimental data are considered. Note that in the calculation the quark wave function is normalized according to $\int d^3\vec{x} u^\dag(\vec x)u(\vec x)=1$. Larger bases have been applied, but the fitted results of the quark wave functions appear the same as the one with the basis of five Sturmian functions.

Shown in the left panel of~\fref{fig:QWF} are the quark radial wave functions $g(r)$ and $f(r)$ with the central values of the expansion coefficients, and in the right panel are the proton charge form factor $G_E^p(Q^2)$ derived with the fitted quark radial wave functions shown in the left panel.
It is seen in right panel of~\fref{fig:QWF} that the theoretical results with the predetermined quark wave functions are well fitted to the experimental data on proton charge form factor up to the squared momentum transfer $Q^2=1.0$ $\textrm{GeV}^2$. The charge radius of proton is derived as $\langle r_E^2\rangle^p=0.77\pm0.10\, {\rm fm^2}$, where the uncertainties arise from the fitting errors of the quark wave functions (the same hereinafter in~\tref{tab:Chargeradius}--\tref{tab:Magneticradius}). The proton charge radius is
consistent with the experimental data $0.76\pm0.02\,{\rm fm^2}$~\cite{PDG12:2012}.

\section{Numerical results and discussion}\label{sec:Results}
The quark wave function has been extracted by fitting the theoretical result of the proton charge form factor to experimental data in the framework of the SU(2) flavor
symmetry. In this section, we study the electromagnetic properties of the baryon octet in the PCQM by applying the predetermined quark wave function. We extend the calculations to the SU(3) flavor symmetry, including kaon
and $\eta$-meson cloud contributions as well. Note that there is no any free parameter in the following numerical calculations on electromagnetic form factors of octet baryons.

\begin{table}
\caption{\label{tab:Chargeradius} Numerical results for the octet baryon mean-square charge radii $\langle r_E^2\rangle^B$ (in units of $\rm fm^2$), where the uncertainties
are from the errors of the quark wave functions. The experimental data are taken from~\cite{PDG12:2012} while the chiral extrapolation estimations of Lattice-QCD results are taken from~\cite{Wang48:2009}.}
\begin{indented}
\lineup
\item[]\begin{tabular}{@{}lccccc}
\br
   \multicolumn{1}{c}{ }&
   \multicolumn{1}{c}{3q}&
   \multicolumn{1}{c}{Meson loops}&
   \multicolumn{1}{c}{\multirow{2}{*}{Total}}&
   \multicolumn{1}{c}{\multirow{2}{*}{Lattice~\cite{Wang48:2009}}}&
   \multicolumn{1}{c}{\multirow{2}{*}{Exp.~\cite{PDG12:2012}}}\\
   \multicolumn{1}{c}{ }&
   \multicolumn{1}{c}{LO+CT}&
   \multicolumn{1}{c}{MC+VC}&
   \multicolumn{1}{c}{ }&
   \multicolumn{1}{c}{ }\\
\mr\rule{0pt}{5pt}
$\langle r_E^2\rangle^p$ & 0.710 & 0.057 & 0.767$\pm0.113$ &0.685(66) &0.76$\pm$0.09\\ \rule{0pt}{15pt}
   $\langle r_E^2\rangle^n$ & 0 & \-0.014 & \-0.014$\pm0.001$ & \-0.158(33) &\-0.116$\pm$0.002\\ \rule{0pt}{15pt}
   $\langle r_E^2\rangle^{\Sigma^+}$ & 0.701 & 0.080 & 0.781$\pm0.108$ & 0.749(72)& ---\\ \rule{0pt}{15pt}
   $\langle r_E^2\rangle^{\Sigma^0}$ & \-0.009 & 0.009 & 0 & --- & ---\\ \rule{0pt}{14pt}
   $\langle r_E^2\rangle^{\Sigma^-}$ & 0.718 & 0.063 & 0.781$\pm0.108$ & 0.657(58) & 0.61$\pm$0.21\\ \rule{0pt}{15pt}
   $\langle r_E^2\rangle^{\Lambda}$ & \-0.009 &  0.009 & 0 & 0.010(9)\0 & ---\\ \rule{0pt}{14pt}
   $\langle r_E^2\rangle^{\Xi^0}$ & \-0.017 & 0.031 & 0.014$\pm0.008$ & 0.082(29)& ---\\ \rule{0pt}{15pt}
   $\langle r_E^2\rangle^{\Xi^-}$ & 0.727 & 0.040 & 0.767$\pm0.113$ & 0.502(47) & ---\\
\br
\end{tabular}
\end{indented}
\end{table}

Listed in~\tref{tab:Chargeradius} are the charge radii squared of the baryon octet. It is found that the uncertainties in the total values of the charge radii squared caused by the fitting errors are estimated around 15\%. The 3q-core (LO and CT diagrams) dominates the charge radii of the charged baryons ($p$, $\Sigma^+$, $\Sigma^-$ and $\Xi^-$), contributing over 90\% to the total values. As shown in~\tref{tab:Chargeradius}, the theoretical $p$ and $\Sigma^-$ charge radii are in good agreement with the experimental values. The work predicts that the charge radii of $\Sigma^+$ and $\Xi^-$ are contributed by a similar pattern based on the SU(3) symmetry, that is, about 90\% from the 3q-core and less than 10\% from the meson cloud contributions (MC and VC diagrams).  The predictions are also closed to the chiral extrapolation estimations of Lattice-QCD values~\cite{Wang48:2009} and the results of relativistic quark model (RQM)~\cite{Ramalho41:2011,Ramalho42:2013}. In~\fref{fig:GEB}, we present the $Q^2$ dependence of the charge form factor of the charged and neutral baryons respectively in the region $Q^2\leq 1$ $\rm GeV^2$, compared with the experimental data \cite{Janssens13:1966,Berger14:1971,Price15:1971,Hanson16:1973,Murphy17:1974,Hohler18:1976,Simon19:1980,Walker20:1994,Eden21:1994,Bruins22:1995,Herberg23:1999,Ostrick24:1999,Passchier25:1999,Golak26:2001,Bermuth27:2003,Madey28:2003,Warren29:2004,Glazier30:2005}. It is seen in the left panel of~\fref{fig:GEB} that the theoretical charge form factor for the proton is consistent with the experimental data, and the charge form factors for hyperons ($\Sigma^+$, $\Sigma^-$ and $\Xi^-$) behave the similar way based on SU(3) symmetry.

\begin{figure}
\begin{center}
\includegraphics[width=7.64cm]{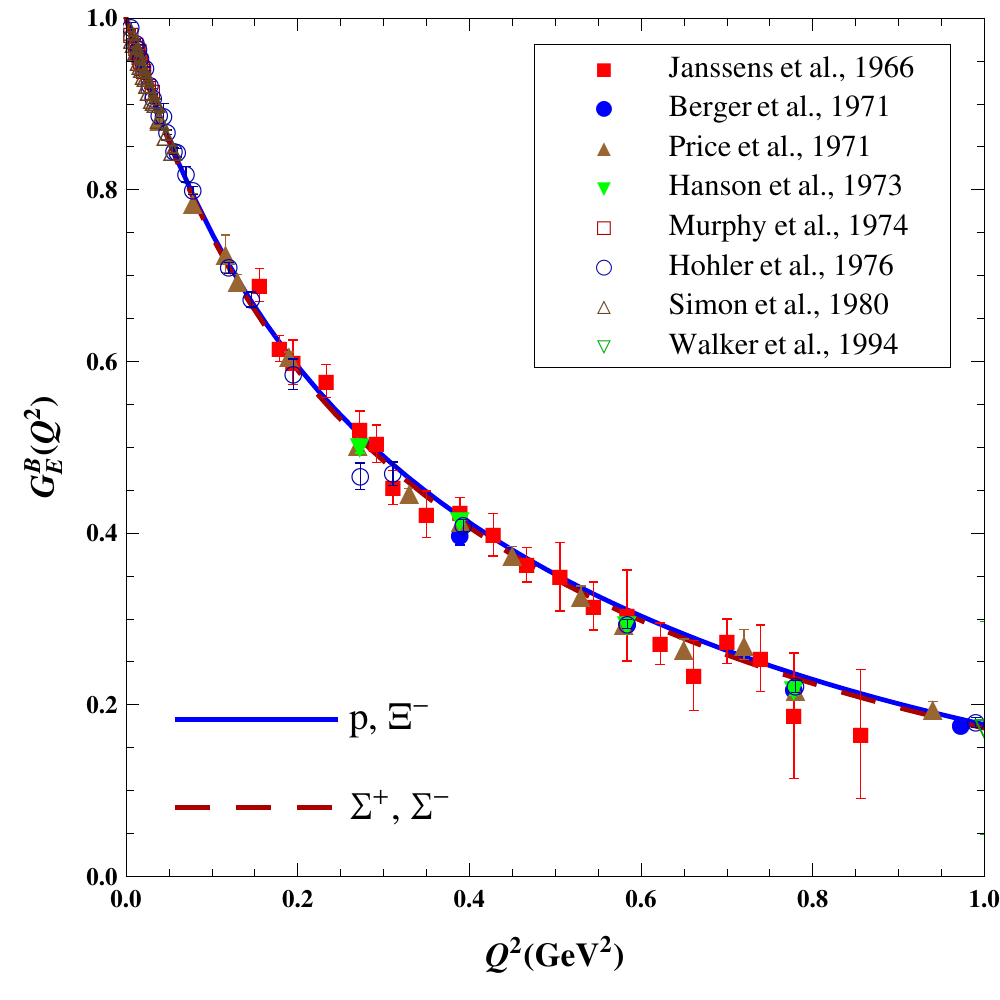}
\includegraphics[width=7.85cm]{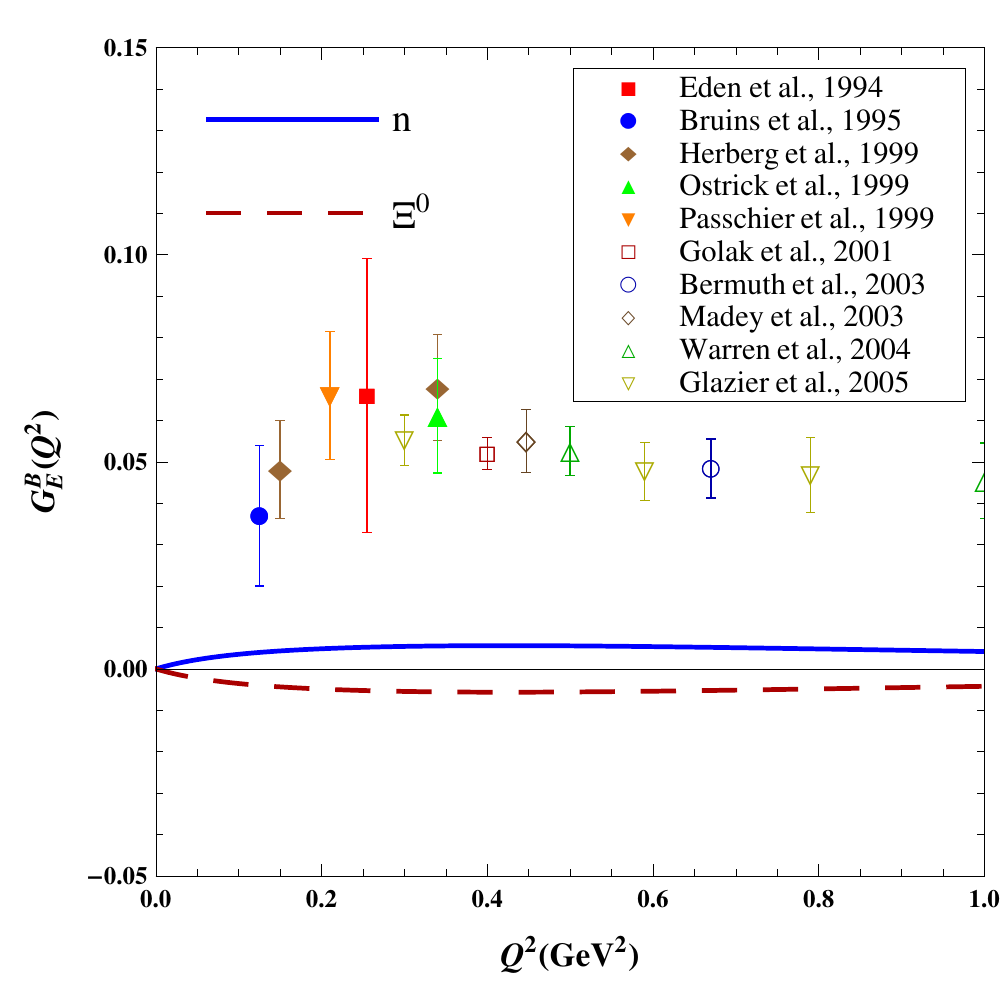}
\end{center}
\caption{\label{fig:GEB}Charge form factors $G_E^B(Q^2)$ of octet baryons. The experimental data on $G_E^p(Q^2)$ taken from~\cite{Janssens13:1966,Berger14:1971,Price15:1971,Hanson16:1973,Murphy17:1974,Hohler18:1976,Simon19:1980,Walker20:1994},
and $G_E^n(Q^2)$ taken from~\cite{Eden21:1994,Bruins22:1995,Herberg23:1999,Ostrick24:1999,Passchier25:1999,Golak26:2001,Bermuth27:2003,Madey28:2003,Warren29:2004,Glazier30:2005}.}
\end{figure}

As shown in~\tref{tab:Chargeradius}, however, the theoretical charge radii of neutral baryons ($n$, $\Sigma^0$, $\Lambda$ and $\Xi^0$) are rather small. The neutron charge radius is much smaller than the experimental data although it is closed to the RQM result of Ref.~\cite{Ramalho41:2011}, while the contributions to the charge radii of $\Sigma^0$ and $\Lambda$ by various diagrams counteract each other to zero. As expected, the work also fails to reproduce the experimental data of the neutron form factor, as shown in the right panel of~\fref{fig:GEB}. The reason might be that the quark propagator is restricted to the ground-state only in our calculation. The meson cloud solely contributes to the neutral baryon charge form factors as the leading-order contribution of the 3q-core vanishes. One may propose that it is necessary to include excited-state quarks to investigate the neutral baryon charge form factors. More discussions and results on the neutron charge radius including the excited quark propagator may be found in~\cite{Lyubovitskij3:2001,Cheedket4:2004}.

\begin{table}
\caption{\label{tab:Magneticmoment} Numerical results for the octet baryon magnetic moments $\mu_B$ (in units of the nucleon magneton $\mu_N$) with chiral mass $m_B=1.039$ \rm GeV, where the uncertainties are from the fitting errors of the quark wave functions. The experimental data are taken from~\cite{PDG12:2012} while the HBXPT extrapolation estimations of Lattice-QCD results are taken from~\cite{Lin43:2009} for $m_\pi=354$ \rm MeV.}
\begin{indented}
\lineup
\item[]\begin{tabular}{@{}lccccc}
\br
   \multicolumn{1}{c}{ }&
   \multicolumn{1}{c}{3q}&
   \multicolumn{1}{c}{Meson loops}&
   \multicolumn{1}{c}{\multirow{2}{*}{Total}}&
   \multicolumn{1}{c}{\multirow{2}{*}{Lattice~\cite{Lin43:2009}}}&
   \multicolumn{1}{c}{\multirow{2}{*}{Exp.~\cite{PDG12:2012}}}\\
   \multicolumn{1}{c}{ }&
   \multicolumn{1}{c}{LO+CT}&
   \multicolumn{1}{c}{MC+VC+MF}&
   \multicolumn{1}{c}{ }&
   \multicolumn{1}{c}{ }\\
\mr \rule{0pt}{5pt}
   $\mu_p$ & 2.290 &0.445 & 2.735$\pm0.121$ & 2.4(2)\0\0 &2.793\\ \rule{0pt}{15pt}
   $\mu_n$ & \-1.527 & \-0.429 & \-1.956$\pm0.103$ & \-1.59(17) &\-1.913\\ \rule{0pt}{15pt}
   $\mu_{\Sigma^+}$ & 2.299 & 0.238 & 2.537$\pm0.201$ & 2.27(16) &2.458$\pm$0.010\\ \rule{0pt}{15pt}
   $\mu_{\Sigma^0}$ & 0.773 & 0.065 & 0.838$\pm0.091$ & ---\0 &---\\ \rule{0pt}{15pt}
   $\mu_{\Sigma^-}$ & \-0.754 & \-0.107 & \-0.861$\pm0.040$ & \-0.88(8)\0 &\-1.160$\pm$0.025\\ \rule{0pt}{15pt}
   $\mu_{\Lambda}$ & \-0.791 & \-0.076 & \-0.867$\pm0.074$ & ---\0 &\-0.613$\pm$0.004\\ \rule{0pt}{15pt}
   $\mu_{\Xi^0}$ & \-1.564 & \-0.126 & \-1.690$\pm0.142$ & \-1.32(4)\0 &\-1.250$\pm$0.014\\ \rule{0pt}{15pt}
   $\mu_{\Xi^-}$ & \-0.800 & \-0.040 & \-0.840$\pm0.087$ & \-0.71(3)\0 &\-0.651$\pm$0.080\\ \rule{0pt}{15pt}
   $\mu_{\Sigma^0\Lambda}$ & \-1.322 & \-0.277 & \-1.599$\pm0.068$ & ---\0 &\-1.610$\pm$0.080\\
\br
\end{tabular}
\end{indented}
\end{table}

\begin{table}[h!]
\caption{\label{tab:Magneticradius} Numerical results for the octet baryon mean-square magnetic radii $\langle r_M^2\rangle^B$ (in units of $\rm fm^2$), where the uncertainties
are from the fitting errors of the quark wave functions.The experimental data are taken from~\cite{PDG12:2012} while the Lattice-QCD values are taken from~\cite{Wang50:2009} for $m_\pi=306$ \rm MeV.}
\begin{indented}
\lineup
\item[]\begin{tabular}{@{}lccccc}
\br
   \multicolumn{1}{c}{ }&
   \multicolumn{1}{c}{3q}&
   \multicolumn{1}{c}{Meson loops}&
   \multicolumn{1}{c}{\multirow{2}{*}{Total}}&
   \multicolumn{1}{c}{\multirow{2}{*}{Lattice~\cite{Wang50:2009}}}&
   \multicolumn{1}{c}{\multirow{2}{*}{Exp.~\cite{PDG12:2012}}}\\
   \multicolumn{1}{c}{ }&
   \multicolumn{1}{c}{LO+CT}&
   \multicolumn{1}{c}{MC+VC+MF}&
   \multicolumn{1}{c}{ }&
   \multicolumn{1}{c}{ }\\
\mr \rule{0pt}{5pt}
   $\langle r_M^2\rangle^p$ & 0.748 & 0.161 & 0.909$\pm0.084$ & 0.470(48) &0.74$\pm$0.10\\ \rule{0pt}{15pt}
   $\langle r_M^2\rangle^n$ & 0.698 & 0.224 & 0.922$\pm0.079$ & 0.478(50) &0.76$\pm$0.02\\ \rule{0pt}{15pt}
   $\langle r_M^2\rangle^{\Sigma^+}$ & 0.810 & 0.075 & 0.885$\pm0.094$ & 0.466(42) &---\\ \rule{0pt}{15pt}
   $\langle r_M^2\rangle^{\Sigma^0}$ & 0.824 & 0.027 & 0.851$\pm0.102$ & 0.432(38) &---\\ \rule{0pt}{15pt}
   $\langle r_M^2\rangle^{\Sigma^-}$ & 0.783 & 0.168 & 0.951$\pm0.083$ & 0.483(49) &---\\ \rule{0pt}{15pt}
   $\langle r_M^2\rangle^{\Lambda}$ & 0.815 & 0.037 & 0.852$\pm0.103$ & 0.347(24) &---\\ \rule{0pt}{15pt}
   $\langle r_M^2\rangle^{\Xi^0}$ & 0.827 & 0.044 & 0.871$\pm0.099$ & 0.384(22) &---\\ \rule{0pt}{15pt}
   $\langle r_M^2\rangle^{\Xi^-}$ & 0.851 & \-0.011 & 0.840$\pm0.109$ & 0.336(18) &---\\ \rule{0pt}{15pt}
   $\langle r_M^2\rangle^{\Sigma^0\Lambda}$ & 0.739 & 0.174 & 0.913$\pm0.083$ & ---\0 &---\\
\br
\end{tabular}
\end{indented}
\end{table}

In our evaluations of the charge form factor of octet baryons we have applied an ansatz that the predetermined quark wave function is the same for u, d, and s quarks.  That is, we work in the SU(3) chiral symmetry limit. Therefore, baryon masses should be restricted to the same order in the calculation of the magnetic moments. We evaluate the magnetic moments with the baryon chiral mass $m_B=1.039$ \rm GeV~\cite{Scherer31:2003}. The numerical results for the magnetic
moments, which are the magnetic form factors in zero-recoil, and the magnetic radii of the octet baryons are given respectively in~\tref{tab:Magneticmoment} and~\tref{tab:Magneticradius}. It is found that the theoretical results for the octet baryon magnetic moments are consistent with the experimental data~\cite{PDG12:2012} and the HBXPT extrapolation estimations of Lattice-QCD values~\cite{Lin43:2009}. Also, our results are in good agreement with the RQM results~\cite{Faessler52:2006,Faessler53:2006,Ramalho41:2011,Gutche55:2012,Ramalho42:2013}.

As listed in~\tref{tab:Magneticradius}, the nucleon magnetic radii are a little bit larger than the experimental data. Our results on light hyperons are in the same order as $\mu_N$ since our calculations are restricted to the SU(3) chiral symmetry. The Lattice-QCD values~\cite{Wang50:2009} listed in~\tref{tab:Magneticradius} are for $m_\pi=306$ \rm MeV, and hence can not be compared directly with our theoretical results owing to the lack of chiral extrapolations. However, we can see that our theoretical results and the Lattice-QCD values are consistent
more or less. For instance, $\langle r_M^2\rangle^{\Sigma^-}$ and $\langle r_M^2\rangle^{\Xi^-}$ take respectively the largest and smallest values for both the works.

As shown in~\tref{tab:Magneticmoment} and~\tref{tab:Magneticradius}, the meson cloud contributes around 20\% to the total values of both the nucleon magnetic moments and radii, while the contributions for hyperons are rather small except for the $\Sigma^-$. It is noted in Ref.~\cite{Cheedket4:2004} that the constants $b_4$, $b_6$ and $b_9$ for hyperons, which are relevant to the $\pi$-meson cloud contribution, are smaller than those for the nucleon. This may indicate that the $\pi$-meson dominates the meson-cloud contribution to the octet baryon magnetic properties. In addition, the uncertainties in the magnetic moments and radii are less than 10\%.

\begin{figure}
\begin{center}
\includegraphics[width=7.78cm]{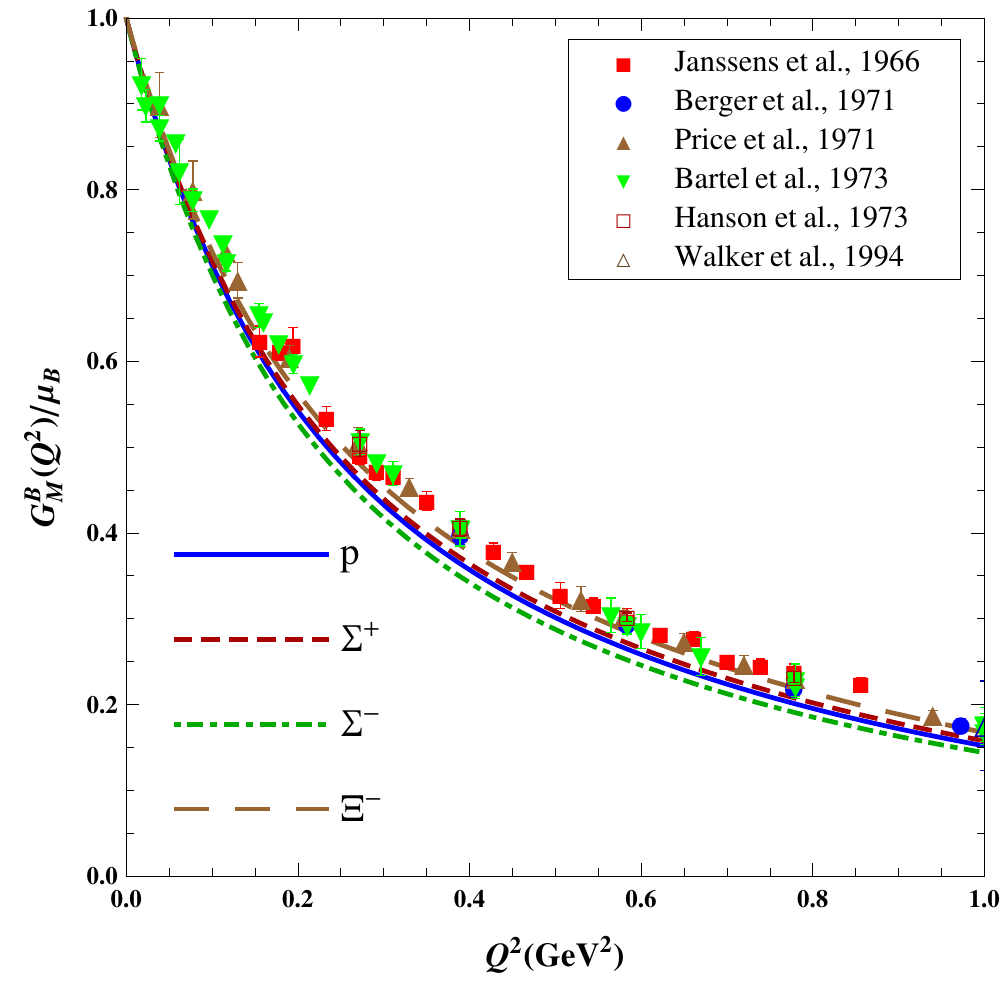}
\includegraphics[width=7.78cm]{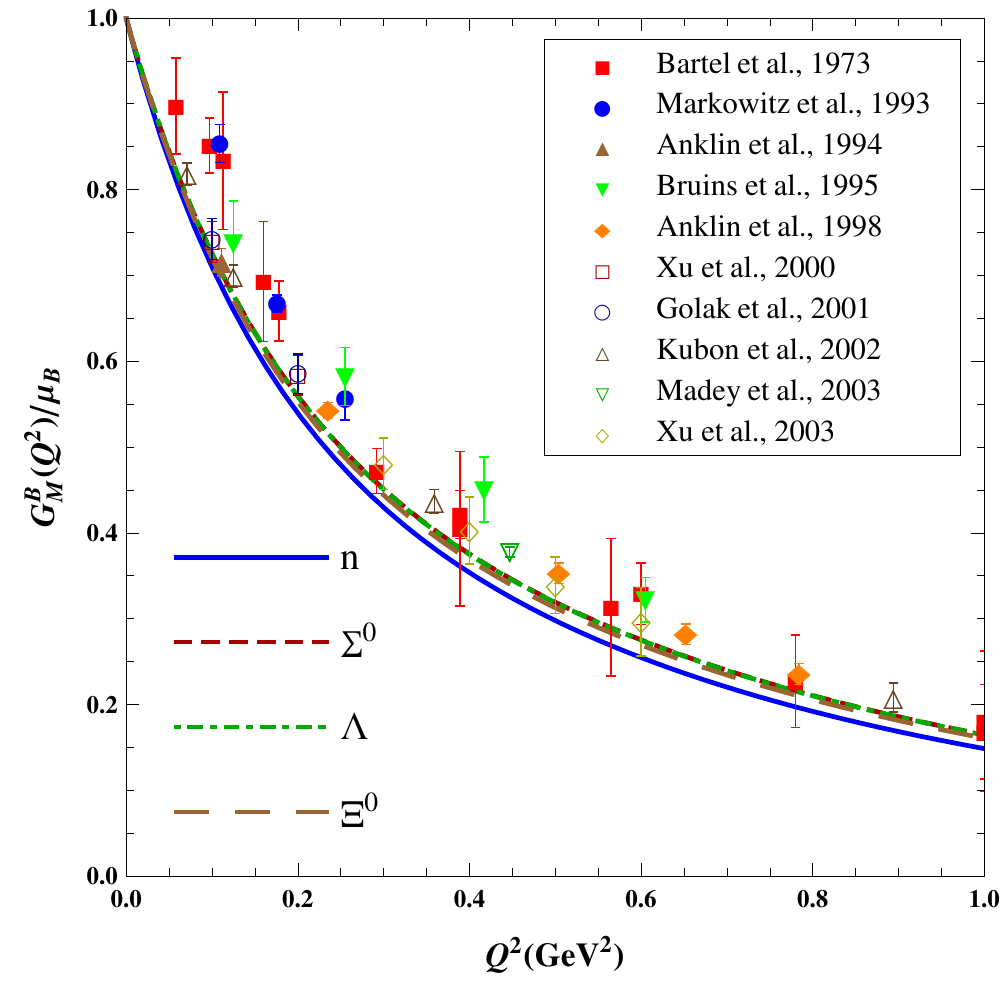}
\end{center}
\caption{\label{fig:GMB}Normalized magnetic form factors $G_M^B(Q^2)/\mu_B$ of octet baryons. The experimental data on $G_M^p(Q^2)/\mu_p$ taken from~\cite{Janssens13:1966,Berger14:1971,Price15:1971,Hanson16:1973,Walker20:1994,Bartel32:1973}, and $G_M^n(Q^2)/\mu_n$ taken from~\cite{Golak26:2001,Bartel32:1973,Markowitz33:1993,Anklin34:1994,Bruins35:1995,Anklin36:1998,Xu37:2000,Kubon38:2002,Madey39:2003,Xu40:2003}.}
\end{figure}

The $Q^2$ dependence of the magnetic form factors for the charged and neutral octet baryons are shown in~\fref{fig:GMB}, which are normalized to one at zero-recoil. We also plot the experimental data on the proton and neutron magnetic form factors in the corresponding figures. It is clear that the nucleon magnetic form factors are fairly consistent with experimental data, and the magnetic form factors for hyperons behave the similar way.

The fact that the $Q^2$ dependence of the theoretical electromagnetic form factors in the region $Q^2\leq1$ $\rm GeV^2$ is consistent with experimental data implies that the predetermined quark wave function is reasonable in the PCQM. We expect that the determined quark wave function is applicable to the evaluation of the axial form factors of baryon octet.

\ack{ This work is supported  by  Suranaree University of Technology (SUT).  XL and KK acknowledge support from SUT scholarship and TRF-CHE (MRG5080374), respectively.}

\end{document}